\documentclass{nature}
\usepackage{graphicx}
\usepackage{color}
\usepackage{calrsfs}

\usepackage{subfigure}
\usepackage{verbatim}
\usepackage{epsfig}% Include figure files
\usepackage{dcolumn}% Align table columns on decimal point
\usepackage{bm}% bold math
\usepackage{amsmath}% 
\usepackage{amssymb} 
\usepackage{enumitem}
\RequirePackage{lineno}

\newcommand{\avg}[1]{\ensuremath{\left< #1 \right>}}

\newcommand{\um}{\ensuremath{\mu \mathrm{m}}}
\newcommand{\invs}{\ensuremath{\mathrm{s^{-1}}}}

\newcommand{\invt}{\ensuremath{\tau_0^{-1}}}

\title{Making soft solids flow: microscopic bursts and conga lines
  in jammed emulsions}

\author{Vishwas V.\,Vasisht$^{1}$, S.\,K.\,Dutta$^{1}$, Emanuela Del
  Gado$^{1}$ \& Daniel L.\,Blair$^{1}$}

\begin{document}

\maketitle

\begin{affiliations}
 \item Department of Physics and Institute for Soft Matter Synthesis
   and Metrology, Georgetown University, 37th and O Streets, N.W.,
   Washington, D.C. 20057, USA
\end{affiliations}

\begin{abstract}
It is well known that jammed soft materials will flow if sheared above
their yield stress -- think mayonnaise spread on bread -- but a
complete microscopic description of this seemingly simple process has
yet to emerge. What remains elusive is a microscopic framework that
explains the macroscopic flow, derived from a 3-D spatially resolved
analysis of the dynamics of the droplets or particles that compose the
soft material. By combining confocal-rheology experiments on
compressed emulsions and numerical simulations, we unravel that the
primary microscopic mechanisms for flow are strongly influenced by the
rate of the imposed deformation. When shearing fast, small coordinated
clusters of droplets move collectively as in a conga line, while at
low rates the flow emerges from bursts of droplet rearrangements,
correlated over large domains. These regions exhibit complex
spatio-temporal correlation patterns that reflect the long range
elasticity embedded in the jammed material. These results identify the
three-dimensional structure of microscopic rearrangements within
sheared soft solids, revealing that the characteristic shape and
dynamics of these structures are strongly determined by the rate of
the external shear.
\end{abstract}

A hallmark of all soft disordered solids is their highly nonlinear
response to the rate of externally applied shear deformations
\cite{liu_nat_1998, vanHecke_JPCM_2009}. Experiments reveal that
granular solids, colloidal glasses and foams exhibit shear
localization that results in a spatially heterogeneous rheological
response \cite{Schall_Science_2007, Seth_NMat_2011,
  goyon_nat_2008,divoux_prl_2010, gopal_JCIS_1999}. The bulk flow
behavior of soft solids is similar to various yield stress fluids,
where the flow curve generically follows an empirical dependence of
the shear stress $\sigma$ on the shear rate $\dot{\gamma}$, $\sigma =
\sigma_Y + K \dot{\gamma}^n$, where $\sigma_Y$ is the yield stress and
$K$ and $n$ are constants specific to the flowing material. For $n\sim
0.5$ the flow curve has a Herschel-Bulkley (HB) form which can be
understood in terms of a non-local constitutive relationship
\cite{hebraud_prl_1998, nicolas_SM_2014}. The underlying assumption is
that, close to yielding, plastic and irreversible rearrangements are
the local mechanism for shear stress dissipation
\cite{argon_AM_1979,falk_Science_2007,Schall_Science_2007,jensen_pre_2014,
  wyart_PRX_2016}. 2-D simulations and mean-field theories reveal that
localised plastic rearrangements occurring within a nearly continuum
elastic background may lead to avalanches \cite{talamali_PRE_2011,
  maloney_pre_2006,
  maloney_JPCM_2008,Martens_PRL_2011,wyart_PRX_2016}. Experiments that
focus on particle scale dynamics also point to the importance of
quantifying local rearrangements and their correlations
\cite{gopal_JCIS_1999,chen_pre_2010, knowlton_softmatt_2014,
  weeks_PRL_2015}.  Interestingly, these local rearrangements are
expected to be rate dependent and could be used, in principle, to
predict the form of the macroscopic flow\cite{lemaitre_prl_2009}. What remains elusive is a microscopic
framework that explains the macroscale rheology derived from a
three-dimensional spatially-resolved analysis of particle level
dynamics.
  
One ideal model soft solid that exhibits universal rheological
nonlinearities is an emulsion, a widely used system for the transport
of highly value-added materials and for the efficient processing of
immiscible fluids \cite{bibette2003emulsion}. By combining
confocal-rheology experiments and molecular dynamics simulations, we
quantitatively describe the droplet scale deformations in a sheared
three-dimensional jammed emulsion. Our combined approach is an
essential step, due in part to microscopic insights provided by
numerical simulations that are often beyond the limits of experimental
resolution. In this work, we find that the elementary rearrangements
reveal unexpected distributions of particle displacements and
localised structures that are spatio-temporally correlated and highly
dependent on the imposed shear rate. The physical picture is that the
flow at high shear rates originates from the highly coordinated motion
of small particle clusters that move collectively in same
direction. Conversely, at low shear rates, the bulk flow emerges from
elementary non-affine rearrangements which take place over much larger
correlated domains that trigger events later in time, determined by
the long-range elastic strain field of the initially solid amorphous
material. The nature itself of the elementary and non-affine type of
rearrangements, which are the primary microscale mechanism for flow,
appears therefore to be dependent on the flow rate. The novel insight
obtained here is a fundamental starting point for the development of
new theories and constitutive models that are needed to design and
control the flow of technologically relevant materials.

\begin{figure}
\includegraphics[width=0.8\textwidth]{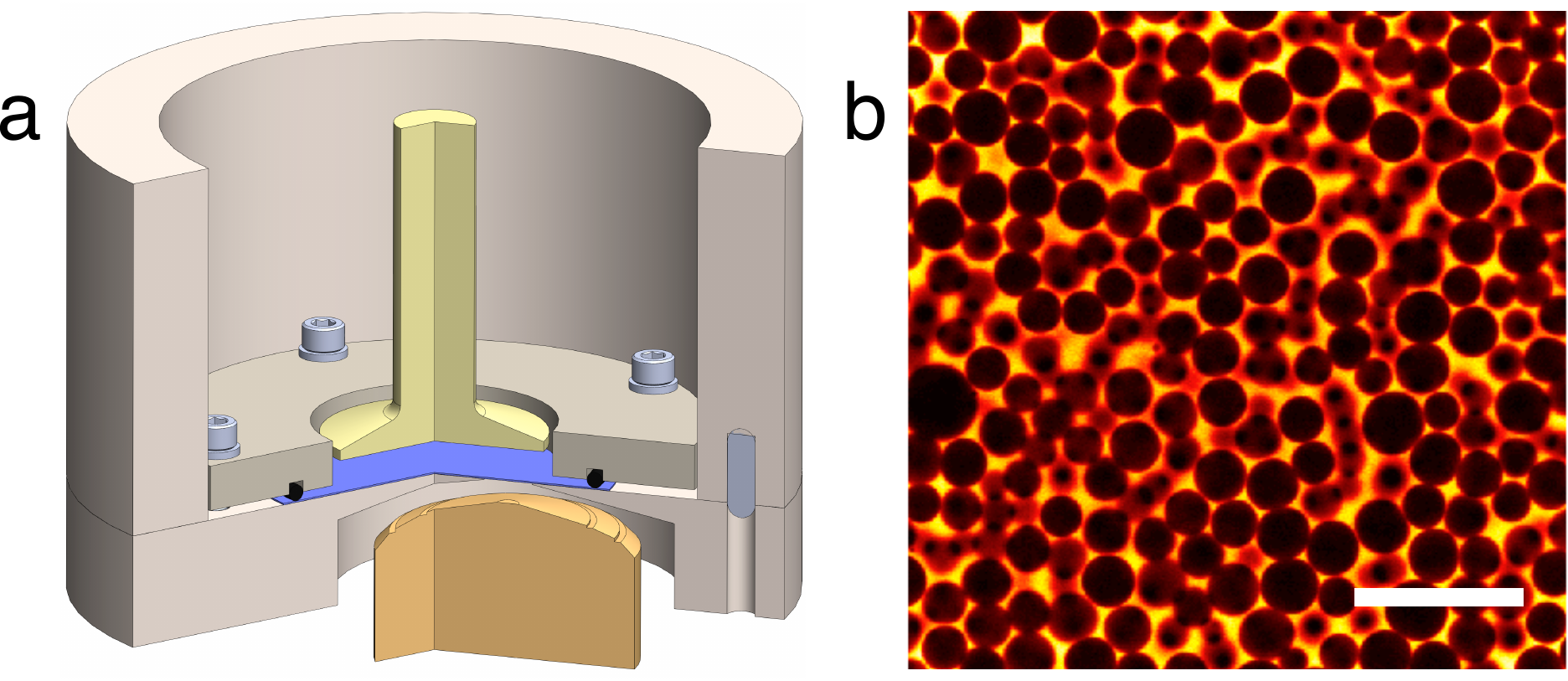}
	\caption{\label{Fig1}{\bf Experimental system details.} {\bf
            a}, Cut-away diagram of the confocal rheometer
          system. Samples are loaded in the gap between the upper tool
          (yellow) and the lower glass plate (blue). {\bf b}, Confocal
          micrograph of the $x,y$-plane showing a cross section of the
          emulsion at volume fraction $\phi=0.70$, scale bar
          represents 20$\mu$m.}
\end{figure}

The experimental system is a direct emulsion of silicone oil droplets
stabilized with the surfactant sodium dodecyl sulfate (SDS) in an
index-matched continuous phase of water, glycerol, and fluorescein
(see Methods).  The droplets, with an average diameter of $6.0\ \mu$m and
polydispersity of 0.15, were compressed above the jamming point with
centrifugation.  Measurements of the system under flow were performed
with a custom stress-controlled rheometer integrated with a laser
scanning confocal microscope.\cite{Dutta_RSI_2013} As shown in
Fig\, \ref{Fig1}a, the rheometer's gap is formed by a glass coverslip,
which provides optical access from below for the microscope, and a
parallel plate tool.  Images were taken at a fixed relative position
between the two instruments, where the local velocity, vorticity, and
gradient axes will be referred to as x, y, and z, respectively.

We measure the influence of shear rate on the local droplet dynamics
by acquiring and analysing time-resolved fluorescence confocal images
while the rheometer simultaneously applies a continuous rotation at a
fixed strain rate $\dot{\gamma}$ (see {\it Methods}). For
$\dot{\gamma} \le 10^{-2}\ \invs$, time-resolved 3-D stacks are
acquired, while at higher shear rates, we acquire 2-D images at
$z$-positions that are equally spaced throughout the rheometer gap
(Fig.\,\ref{Fig1}b). In the case of 2-D imaging, where particle
locating and tracking is not possible, time-resolved spatial
cross-correlations are calculated between pairs of consecutive images
at a given $z$-position to quantify the spatially-resolved
particle-scale velocities $v(x,y)|_z$, which are compared against
direct particle tracking. The averaged velocities are calculated as a
function of $z$ to determine the velocity profiles at each
$\dot\gamma$ and obtain the local shear rate $\dot{\gamma}_{l}$; this
analysis is performed for two different volume fractions and we
present data for $\phi=0.70$ (see {\em Supplementary Information}).

\begin{figure}
        \includegraphics[width=0.8\textwidth]{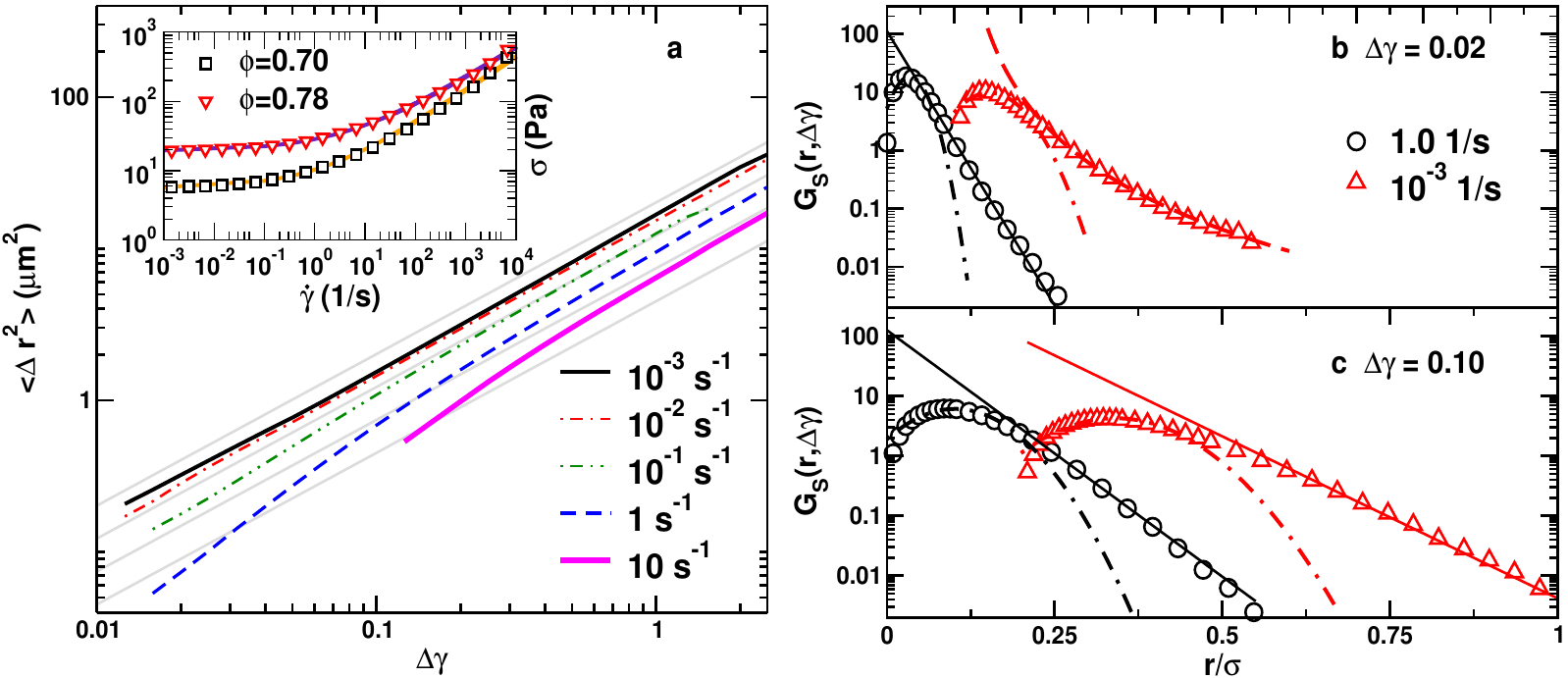}
	\caption{ \label{Fig2}{\bf Flow curve and shear frame
            displacement from experiments.} {\bf a}, The shear-frame
          MSD measured in the $x,y$-plane for shear rates
          $10^{-3}\ \invs$, $10^{-2}\ \invs$, $10^{-1}\ \invs$,
          $1.0\ \invs$ and $10.0\ \invs$. The gray solid lines are
          indicative of diffusive behaviour. (inset) The shear stress
          $\sigma$ as a function of the shear rate $\dot{\gamma}$
          measured at $\phi=0.70$ and $\phi=0.78$ and the solid lines
          represent Herschel-Bulkely fit, with flow indices of $n =
          0.51$ and $n=0.45$.  The linear viscoelastic modulus of
          these emulsions is $G^{\prime} = {180,360}$ Pa for $\phi =
          {0.70, 0.78}$. The experimental van Hove correlation
          function $G_S(r,\Delta \gamma)$ is computed at $\phi=0.70$
          for two different strain windows {\bf b}, $\Delta
          \gamma=0.02$ and {\bf c}, $\Delta \gamma=0.1$ for two
          different shear rates $1.0\ \invs$ and $10^{-3}\ \invs$. For
          the purpose of clarity we have shifted the abscissa of the
          distribution function by a constant factor of $0.2$ for
          $10^{-3}\ \invs$. The dot-dashed line is the best fit with a
          Gaussian, the solid line represents the best fit with an
          exponential, whereas the dashed line is the best fit with a
          power law.}
\end{figure}

The flow curves for our emulsions exhibit HB rheology, with $n
\approxeq 0.51$ for $\phi=0.70$ and a corresponding dependence on
$\sigma_y$ that increases with $\phi$ (see Fig.\,\ref{Fig2} {\em
  inset}). After removing the background flow in the confocal images,
we obtain a vector displacement field, with components $\Delta x$ and
$\Delta y$, of local fluctuations in the shear frame. In
Fig.\,\ref{Fig2}a, we plot the mean square displacement (MSD)
$\avg{\Delta r^2} = \avg{\Delta x^2 + \Delta y^2}$, as a function of
the accumulated strain $\Delta \gamma = \dot{\gamma}_l \Delta t$, over
a wide range of applied shear rates. At the highest shear rates
($\dot{\gamma} \ge 1 \, \invs)$, the system exhibits super-diffusive
behavior at small $\Delta \gamma$, followed by a diffusive regime,
whereas at lower strain rates the $\avg{\Delta r^2} \propto \Delta
\gamma$ over the entire range. The distributions of droplet
displacements for each $\Delta \gamma$ are characterised by
non-Gaussian statistics for all strains. The self-part of the van Hove
correlation function computed for strain windows $\Delta \gamma=0.02$
and $\Delta \gamma=0.10$, shown in Figs.\,\ref{Fig2}b,c indicates
that, in addition to the overall non-Gaussianity for all $\Delta
\gamma$ and each $\dot\gamma$, another distinctive feature arises. At
large $\Delta \gamma$, the distributions exhibit exponential tails for
all $\dot\gamma$, whereas at small $\Delta \gamma$ we can clearly
distinguish a power-law tail for the distributions measured at low
$\dot\gamma$. These observations indicate that the underlying
microscopic motion is far from the simple picture of shear induced
diffusive motion.  Moreover, the data also highlight distinctive
features that depend strongly on the imposed shear
rate. Interestingly, the exponential tails are reminiscent of
correlated and heterogeneous dynamics in supercooled liquids
\cite{chaudhuri_prl_2007}. The power law tails, on the other hand,
suggest a closer connection with plastic rearrangements taking place
in an elastic background, and that the elasticity of the jammed solid
plays a significant role in the limit of quasi-static shear
\cite{picard_EPJE_2004, lemaitre_prl_2009, chattoraj_pre_2011}.

One of the major challenges associated with experiments of jammed
matter is that the characteristic length and time scales rapidly
exceed the resolution of fast confocal microscopy methods. To overcome
this, we employ complementary three-dimensional numerical simulations
of polydisperse spheres that interact through a short ranged repulsion
and are subjected to shear deformations. Our simulations are guided by
the details of the experimental system where the emulsion droplets are
generally stable to both ripening and coalescence and are nearly
athermal in the range of $\phi$ explored. We utilise the WCA form of
the truncated Lennard-Jones potential, defined as $U(r) = 4\epsilon \left [(\sigma_{ij}/r_{ij})^{12} - (\sigma_{ij}/r_{ij})^{6} \right ] + \epsilon$, for
$r_{ij} \le 2^{1/6} \sigma_{ij}$, else $U(r_{ij})=0$
\cite{WCA_JCP_1971}. $\sigma_{ij} = (\sigma_i+\sigma_j)/2$ defines the
distance between the center of particles $i$ (with diameter
$\sigma_i$) and $j$ (with diameter $\sigma_j$) at contact, and
$\epsilon$ defines the energy unit, independent of particle
diameter. We choose $\sigma_{i}$ from a Gaussian distribution with
a variance of 10\%. The numerical sample consist of 97556 particles at
$\phi \sim 70\%$, which corresponds to a cubic simulation box dimension
of $l=42\sigma$.  The finite shear rate simulations were carried out
solving, with the Lees-Edwards boundary conditions, the Langevin
equation of motion
\begin{equation}
        m \frac{d^2\vec{r}_i}{dt^2} = -\zeta \left
        (\frac{d\vec{r}_i}{dt} - \dot{\gamma} z_i \vec{e}_x \right ) -
        \triangledown_{\vec{r}_i} U,
\end{equation}
where $\dot{\gamma}$ is the applied shear rate, $m$ is the particle
mass, $\vec{r}_i$ is the position vector of the particle $i$, $z_i$ is
the $z$-coordinate of particle $i$, $\vec{e}_x$ is the unit vector
along the $x$-axis, $\zeta$ is the damping coefficient and $m/\zeta =
2.0$. $\triangledown_{\vec{r}_i} U$ defines the force to interactions
between the particle. $\dot\gamma$ is expressed in units of $\invt$,
where $\tau_0 = \sigma \sqrt{m/\epsilon}$ is the time unit. All data
presented are in the steady state flowing regime (see {\it Methods}).

\begin{figure}
       \includegraphics[width=0.8\textwidth]{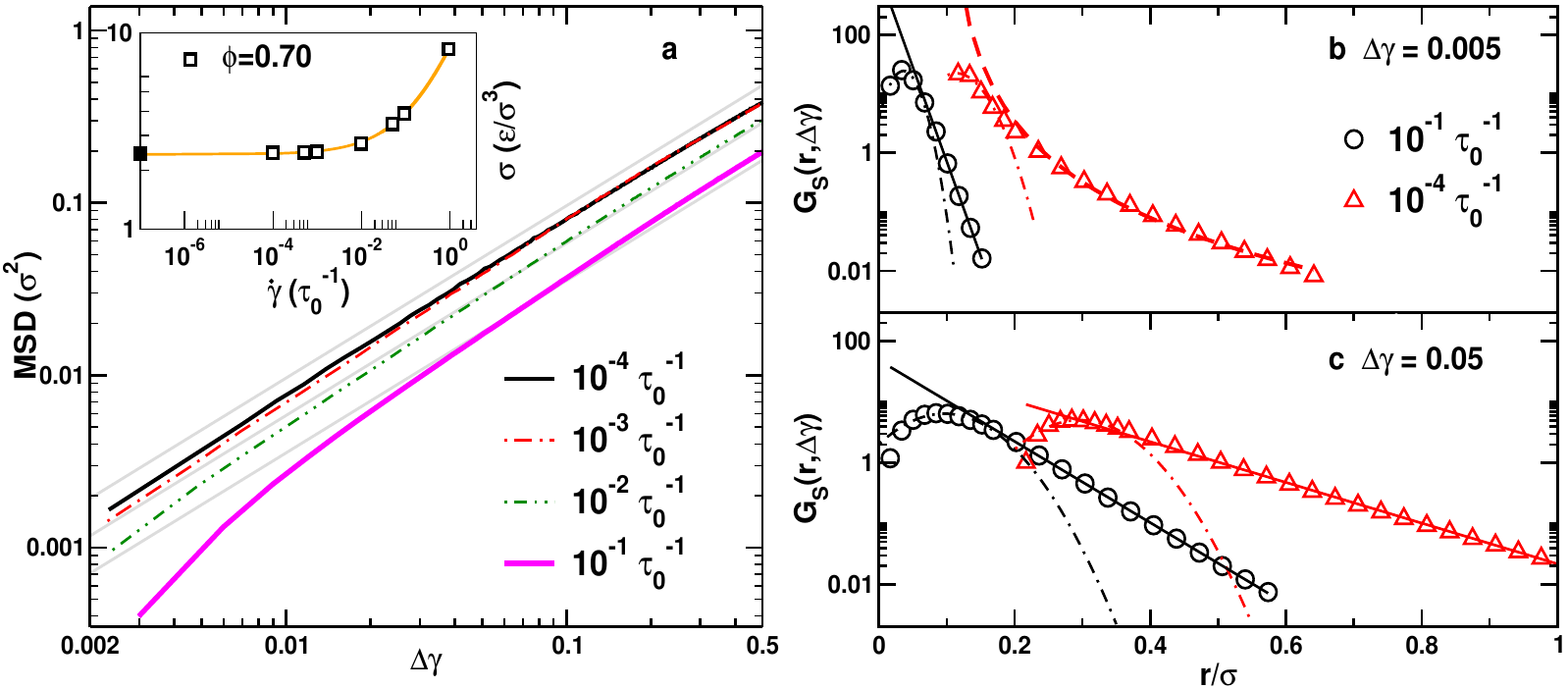}
       \caption{\label{Fig3}{\bf Flow curve and shear frame
           displacement from simulations.} {\bf a}, The non-affine MSD
         computed in the $x,y$-plane for shear rates $10^{-4}
         \ \invt$, $10^{-3} \ \invt$, $10^{-2}\ \invt$ and
         $10^{-1}\ \invt$ for simulations. (inset) Simulation flow
         curve. The shear stress $\sigma$ as a function of the shear
         rate $\dot{\gamma}$ measured at volume fraction $\phi=0.70$
         with solid lines showing a Herschel-Bulkely fit, with a flow
         index of $0.6$. The solid square data obtained from the
         quasi-static simulations which corresponds to a zero shear
         rate condition.  The $G_S(r,\Delta \gamma)$ computed for {\bf
           b}, $\Delta \gamma=0.005$ and {\bf c}, $\Delta
         \gamma=0.005$ for shear rates $\dot{\gamma}=10^{-1}\ \invt$,
         and $\dot{\gamma}=10^{-4}\ \invt$. For the purpose of clarity
         we have shifted the abscissa of the distribution function by
         a constant factor of $0.2$ for $10^{-4}\ \invt$. The
         dot-dashed line is the best fit with a Gaussian, whereas the
         solid and dashed lines correspond respectively to the best
         fit with an exponential and power law.}
\end{figure}

The flow curve obtained from the simulations at $\phi=0.70$ follows
the HB form with a flow index $n\simeq 0.6$ (Fig.\,\ref{Fig3}a {\em
  inset}). We note that the shear stress at $\dot{\gamma}=10^{-4}
\invt$ does not vary much in comparison with the {\it quasi-static}
simulations data (see Fig.\,\ref{Fig3}a {\em inset}),
which represents the zero shear rate limit.

To strengthen the connection between our experiments and simulations,
we perform an identical analysis of simulated displacement statistics
by computing the MSD and the self-part of the van Hove correlation
functions (see Fig.\,\ref{Fig3}). To calculate the non-affine
displacements and correlations, we divide the simulated emulsion into
slices of thickness $2\sigma$ along the $z$-direction and average the
data over each slice. Each slice contains approximately 4000 particles
(see {\em Supplementary Information}). Remarkably, for conditions
equivalent to those shown in Fig.\,\ref{Fig2}, the simulations capture
all of the essential features found in the experimental results. Most
notably, for small $\Delta \gamma$ the MSD obtained in the simulations
exhibits a crossover from ballistic to diffusive motion with
increasing $\dot{\gamma}$, while the van Hove functions exhibit the
same change from exponential to power law behavior with decreasing
$\dot{\gamma}$. The striking similarities in these essential
microscopic metrics provide us enhanced confidence that other
particle-scale features gleaned from our simulations will allow us to
draw physically relevant conclusions about these materials.

The microscopic dynamics are explored in the numerical simulations
through an analysis of the spatio-temporal correlations of the droplet
displacements. We measure the displacement correlations $S_U$, and the
correlations in the deviation from the average displacement $S_{\delta
  U}$, where $\vec{U} = U_x \hat{i} + U_y \hat{j}$ is the displacement
vector of a droplet in the $x,y$-plane and $\delta U = \left | U -
\left< U \right > \right |$ is the average droplet {\it mobility},
(see {\em Methods}). Completely correlated (anti-correlated)
displacements give $S_U=1.0$ ($S_U=-1.0$). $S_{\delta U}=1.0$ if the
fluctuations in displacements are perfectly correlated and both
functions go to zero for uncorrelated motion. $S_{U}$, computed at
fixed $\Delta \gamma$ as a function of distance $r$ from a reference
droplet, can be decomposed into longitudinal $S_{U}^L$ and transverse
$S_{U}^T$ components \cite{weeks_JPCM_2007} (see {\it Methods}).

\begin{figure}
  \includegraphics[width=0.8\textwidth]{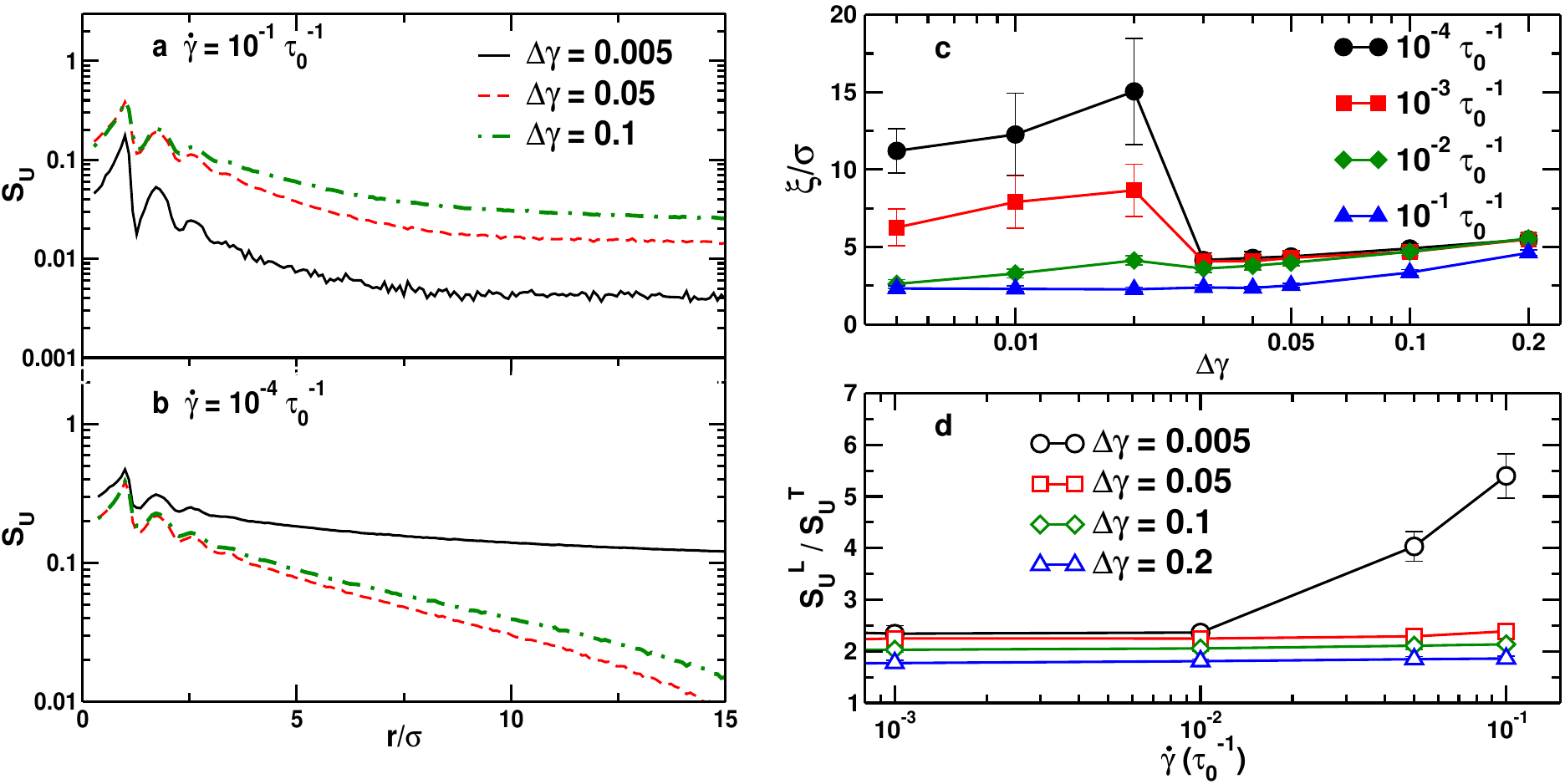}
  \caption{\label{Fig4}{\bf Displacement correlation function from
      simulations.} {\bf a}, At high shear rate $\dot\gamma =
    10^{-1}\ \invt$ and {\bf b}, low shear rate
    $\dot\gamma=10^{-4}\ \invt$, the spatial correlation function
    $S_U$ associated with droplet displacements is shown for shear
    strains $\Delta \gamma = 0.005$ (black), $\Delta \gamma = 0.05$
    (red) and $\Delta \gamma = 0.1$ (green). {\bf c}, the correlation
    length obtained from $S_U$ as a function shear strain, for shear
    rates $10^{-4}\ \invt$, $10^{-3}\ \invt$, $10^{-2}\ \invt$ and
    $10^{-1}\ \invt$. {\bf d}, the ratio of longitudinal to transverse
    displacement correlation function measured at a distance around $1
    \sigma$, for accumulated strains $0.005$, $0.05$, $0.1$ and $0.2$
    as a function of shear rate.}
	
\end{figure}

We compute $S_{U}(\Delta\gamma,\dot\gamma)$ and $S_{\delta U}
(\Delta\gamma, \dot\gamma)$ at $\dot\gamma= 10^{-1}$ and $10^{-4}
\invt$ over a range of $\Delta \gamma$. In each case, we extract a
correlation length $\xi$ by fitting the data with an exponential
function and averaging the fit parameters over the $z$-direction. At
high shear rates, and for all strain values, both correlations
functions decay exponentially. For clarity, only $S_{U}$ is shown in
Figs.\,\ref{Fig4}a and b see {\em Supplementary Information} for
$S_{\delta U}$.  The corresponding correlation lengths are shown in
Fig.\,\ref{Fig4}c.

Interestingly $S_U^L$ is the dominant correlation component within the
super-diffusive regime of the MSD indicating that the non-affine
motion in this regime is strongly correlated in direction (see
Fig.\,\ref{Fig4}d) \cite{weeks_JPCM_2007}. The correlation length
$\xi_{S_U}$ grows with an increase of $\Delta\gamma$ from $2 \sigma$
to $5 \sigma$ in the same strain interval
(Fig.\,\ref{Fig4}c). Overall, these results suggest that at high shear
rates and small strains, corresponding to the super-diffusive regime,
the motion is localised to small groups of droplets that move in
linear clusters whose size increases with $\Delta\gamma$; this
behavior also carries over into the diffusive regime, where $S_U^T
\simeq S_U^L$ (see Fig.\,\ref{Fig4}d) and persists down to shear rates
$\dot\gamma \to 10^{-2} \invt$. Elementary flow events of this type
are reminiscent of the {\em string-like} motion in the dynamical
heterogeneities of supercooled liquids \cite{donati_PRL_1998} (see
Supplemenatary Movies M3-M8).

When $\dot\gamma < 10^{-2}\ \invt$ (Fig.\,\ref{Fig4}b), there is a
qualitative change in the nature of the elementary rearrangement
events. At small $\Delta\gamma$, both $S_U$ and $S_{\delta U}$ show an
exponential initial decorrelation that is followed by a power-law
decay. The microscopic picture that emerges from these data is that at
low $\dot \gamma$, the droplet rearrangements underlying the
macroscopic flow may resemble plastic-like yielding regions embedded
in an elastic medium. The picture is confirmed by the observed power
law tails in the self part of the van Hove distributions (see
Figs.\,\ref{Fig2} and \ref{Fig3}). At these low deformation rates,
$S_{\delta U}$ is significantly larger than $S_{U}$ at a fixed length
scale and the related correlation length is significantly larger,
$\xi \simeq 10 \sigma$, roughly $25\%$ of the total system size
(Fig.\,\ref{Fig4}c). These observations quantitatively indicate that
the droplet rearrangements involve directionally cooperative mobile
clusters that are embedded in an immobile elastic background. The
origin of the large error bars in the correlation length at these low
rates point to large fluctuations in the instantaneous velocity
profiles, revealing a flow behavior that is both inhomogeneous and
intermittent. At larger $\Delta\gamma$, a larger percentage of
droplets become mobile and we measure a dramatic drop in the
directional correlation length $\xi_{S_U}$, whereas $S_{\delta U}$
continues to show long range correlations. These features clearly
indicate that droplet dynamics are highly dependent on shear rate and
that local rearrangements have a fundamental different structure at
low $\Delta\gamma$. The analysis of the spatio-temporal correlations
indicate that their range and intensity change qualitatively with the
rate.  The two results together suggest that the change in the
spatio-temporal correlations may be at the origin of (or must be
coupled to) the differences in the elementary flow events when the
material is sheared at different rates. Elaborating further, the
differences and dependence of the microscopic non-affine dynamics with
the imposed shear rate are an essential component of the non-linear
dependence on the rate of the flow curve.

\begin{figure}
	\includegraphics[width=0.8\textwidth]{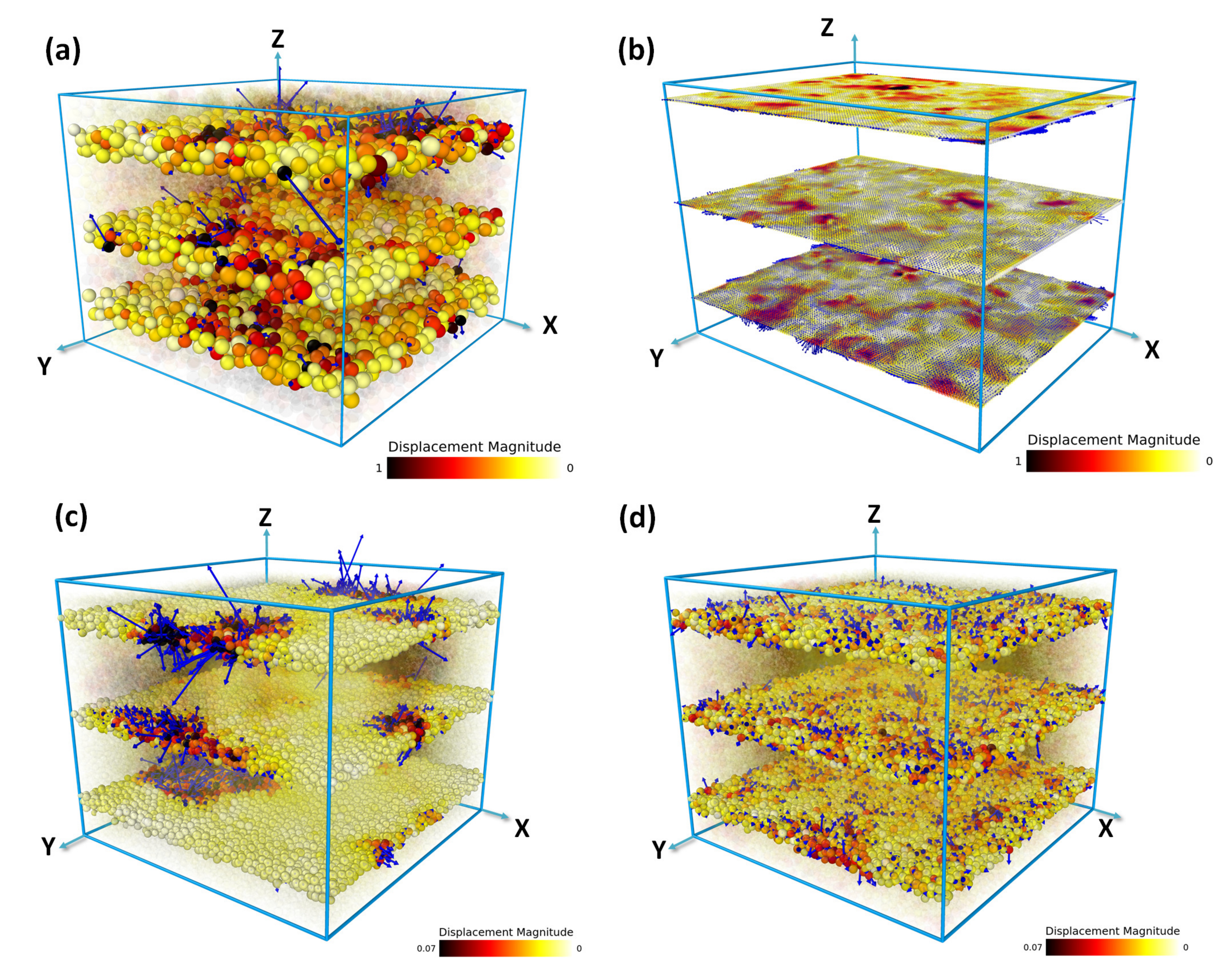}
	 \caption{\label{Fig5}{\bf Displacement maps for experiments
             and simulations.} Experimental trajectory snapshots for
           {\bf a}, $\dot{\gamma}=3\times10^{-3}\ \invs$ and {\bf b},
           $\dot{\gamma}=1.0\ \invs$. For the low shear rate, the
           snapshot is obtained at strain interval of $\Delta \gamma
           \approx 0.02$ (bottom plane), $\Delta \gamma \approx 0.01$
           (middle plane) and $\Delta \gamma \approx 0.001$ (top
           plane). For the high shear rate, the accumulated strain
           intervals are $\Delta \gamma \approx 0.025$ (bottom plane),
           $\Delta \gamma \approx 0.015$ (middle plane) and $\Delta
           \gamma \approx 0.02$ (top plane). The rendered particle
           positions are approximately taken from the middle third of
           the rheometer gap. Simulation trajectory snapshots for
           slow,{\bf c} $\dot{\gamma}=10^{-4}\invt$, and fast {\bf d}
           $\dot{\gamma}=10^{-2}\invt$,at $\Delta \gamma=0.001$. Only
           three different layers in the 3D system are highlighted for
           the purpose of clarity. The colour scheme in all snapshots
           represent non-affine displacement magnitude. The arrows
           represent the non-affine displacement vector. To highlight
           the localisation of displacement vectors, the length of the
           vectors are amplified by 5 times in experimental snapshots
           and by 40 times in simulation snapshots. Renderings are
           made using freely available software
           \cite{stukowski_MSE_2010}.}
\end{figure}

The non-affine displacement maps obtained at fixed shear rates
qualitatively support our conclusions. Representative maps for
experiments and simulations are shown in Fig.\,\ref{Fig5}, where the
colouring is based on the magnitude of non-affine displacements. At
high $\dot\gamma$, we observe relatively small displacements that are
rather spatially homogeneous (see Figs.\,\ref{Fig5}a,c), while at
lower values, the displacements become highly localised, {\em bursty}
and heterogeneous (see Figs.\,\ref{Fig5}b,d and movies M1 and M2 in
the {\em Supplementary Information)}.

Our results provide a first glimpse into the non-equilibrium flow
properties of technologically significant materials by coupling the
microscopic dynamics to the macroscopic flow behavior. We unravel
particle scale rearrangements and spatio-temporal correlations that
are directly connected to the imposed rate of shear deformation. The
steady-state response of unjammed emulsions, defined by the bulk
rheological flow curve of the material, is associated with changes in
the nature of the elementary flow events that are governed by the
onset of microscopic flow.  When the flow is parameterised by the
local accumulated strain, two distinct and rate dependent microscopic
mechanisms are observed as the initiators of stress relaxation and
yielding. Correlated string-like motion, a signature of dynamical
heterogeneities in liquids, is found in the flowing state well above
the yield stress. In contrast, near the yield stress, the primary
mechanism for stress relaxation is localised plastic bursts embedded
within an elastic continuum. These bursts exhibit long range
spatio-temporal correlations and are strongly reminiscent of the local
elastically interacting rearrangements that underlie plasticity in
amorphous solids, referred to as shear transformations. The distinct
and rate dependent nature of these elementary excitations is a
fundamental starting point for the development of microscopic theories
and constitutive models. We provide a quantitative framework for the
differences in the spatio-temporal patterns of the correlations in the
microscopic dynamics, which have themselves a distinct and rate
dependent structure. The new insight gained is a first 3-D microscopic
picture of the spatial correlations in the flow
\cite{goyon_nat_2008,mehdi_PRL_2013,bonn-2015} and of the non-linear
rheology of soft jammed materials.
 
{\it Acknowledgements}. D.L.B.\, and S.K.D.\, acknowledge the support
of the John F. Templeton Foundation Grant No. 57392 and the National
Science Foundation Grant DMR-0847490. They also thank E.D.\,Knowlton,
A.\, Mbi, P.\, Kumar, and R.\,C.\, Arevalo for their experimental and
technical assistance. V.V.V.\, and E.D.G.\, acknowledge the support of
the Swiss National Science Foundation (Grants No. PP00P2\_150738) and
ETH Zurich. All authors thank Georgetown University for their support.

{\it Author Contributions.}  The experiments were initiated, designed
and analysed by D.L.B and S.K.D. Simulations were conceived, designed
and analysed by E.D.G. and V.V.V. The experiments and simulations
were performed by S.D.K. and V.V.V., respectively. All authors
discussed data and aided in preparing the manuscript.

\begin{methods}
\subsection{Sample preparation and data acquisition in experiments.}
\label{SSample}

A highly polydisperse emulsion at oil volume fraction $\phi = 0.7$ was
initially formed by mixing silicone oil (50 cSt, from Clearco Products
Co.)  with a 10 mM solution of sodium dodecyl sulfate (SDS) in
water/glycerol (9:1, by volume) with a laboratory disperser (IKA Ultra
Turrax T25 basic).  We fractionated the sample using an iterative
creaming technique that takes advantage of the oil having a lower mass
density than the continuous phase \cite{Bibette_JCIS_1991}. Rather
than relying on depletion from excess surfactant to flocculate
droplets of a certain size as in the published technique, we found
that the relatively large droplets of interest creamed in a reasonable
time by themselves, so the process was carried out at 10 mM SDS.
After 9 steps, the selected droplets had an average diameter of
$6.0\ \mu$m with a polydispersity of 0.15, as determined by confocal
imaging.  For these experiments, large droplets were chosen so that
small displacements (relative to the average droplet diameter) could
be resolved.  However, even larger droplets would require too large a gap
(to avoid boundary effects) and would be more fragile (due to their low
Laplace pressure).

In preparation for imaging, we index matched the emulsion with
glycerol to obtain a transparent sample.  The resulting continuous
phase was dyed with fluorescein sodium salt at a 3 mM concentration.
The SDS concentration was boosted to 50 mM, which we found increased
the stability of the emulsion, particularly under shear with a high
glycerol concentration.  Even at the elevated surfactant levels, there
was significant droplet coalescence for shear rates $\dot{\gamma} >
10\ \invs$.  While 50 mM is well above the critical micelle
concentration, we could not detect a dependence of the rheology on the
SDS concentration, and thus suspect any depletion force had a minimal
effect.  The samples were compressed to the desired volume fraction
$\phi$ with centrifugation, with the value of $\phi$ determined by
boiling the continuous phase off.

The sample is placed on a glass coverslip at the bottom of a metal cup
that is rigidly mounted to the rheometer.  Optical access for the
microscope is provided from below by a small hole on the underside of
the cup. The flow experiments were performed using a parallel plate
rheometer tool with a diameter $d = 25\ \um$.  The rheometer gap was
set to a final value of $h = 100\ \um$ by lowering the tool slowly
over several minutes, while rotating at 0.1 rpm.  This kept the normal
force low, as monitored by the rheometer, and minimised damage to the
sample. On the bottom plate coverslip, we lithographically defined a
square grid of $10\ \um$ wide, $5\ \um$ high posts (spaced by
$10\ \um$) made from SU-8, a negative photoresist, to minimise surface
slip.  The posts were oriented in a consistent way with respect to the
imaging axes.  A similar slide was affixed to the rheometer tool for
the same purpose.  While some slip was present at high shear rates,
determination of the local strain rate $\dot{\gamma}_l$, for example,
came from direct optical measurements taken over the full gap, without
regard for the nominal rheometer settings.

Images were acquired with a $63 \times$, 1.2 numerical aperture,
water-immersion objective. Water was chosen over oil as the immersion
fluid, as it produces smaller deflections of the coverslip with a
better index-match to the sample. The objective was positioned at a
radial distance of $d / 3$ from the central rheometer axis, a location
where the local strain is equal to the plate-averaged strain.

When performing the flow measurements, the rheometer was set to a
particular shear rate $\dot{\gamma}_0$, as referenced to the edge of
the tool.  In presenting the microscope data, we will refer to the
shear rate at the radius where the images were acquired, $\dot{\gamma}
= 2 \dot{\gamma}_0 / 3$.  After setting $\dot{\gamma}_0$, the imaging
was not initiated until the shear stress stabilised.  We also
attempted to avoid imaging volumes where large coalesced drops were
located.  For $\dot{\gamma} = 10^{-3}\ \invs$ and below, it was
possible to take 3-D stacks with the desired interval in time.  At
$\dot{\gamma} = 10^{-2}\ \invs$, it was only possible to acquire 3-D
stacks that covered a fraction of the gap, so multiple stacks were
acquired to cover the full sample.  For higher rates, we acquired a
rapid set of 2-D slices at fixed $z$, incrementing the $z$ position
between sets.  The time spacing between frames was set to $0.03 /
\dot{\gamma}$, which corresponds to a motion of $3\ \um$ of the tool
at the imaging location.  The image size was $145\ \um \times
145\ \um$ with a resolution of $512 \times 512$.  The exception is for
the data at $\dot{\gamma} = 10\ \invs$, where, due to limitations of
the microscope, the time spacing was set to $0.17 / \dot{\gamma}$ with
the resolution reduced to $512 \times 192$.  In all cases, we acquired
at least 150 frames at each setting of $z$.

\subsection{Sample preparation, shearing protocol, steady state
  analysis in simulations.}  The initial sample on which rheological
studies are carried out is prepared using the following protocol. A
FCC crystal containing $97556$ LJ particles at a desired volume
fraction of $70\%$ (with a system dimension of
$l_x=l_y=l_z=42.17\sigma$) and a continuous polydispersity of $10\%$
is initially melted and equilibrated for 50000 molecular dynamics (MD)
steps, with a timestep of $\Delta t = 0.001$, at a high temperature of
$T=5.0 \epsilon/k_B$. In the melt we do not find any signature of
crystallinity from the FCC initialisation. For the purpose of
measuring crystallinity, we measure the local orientational ordering
$q_6$\cite{steinhardt_PRB_1983}. Note that the orientational ordering
computation requires calculating the nearest neighbours for a
reference particle. For this purpose we define all the particles
within the first Voronoi shell as neighbours. The equilibrated melt is
subjected to a temperature quench, wherein the system at $T=5.0
\epsilon/k_B$ is lowered to a temperature of $T=0.001 \epsilon/k_B$ at
cooling rates of $5\times 10^{-4}$. We then perform conjugate gradient
minimisation to take the system to the zero temperature limit. In this
configuration, we analyse the local orientational ordering and do not
find signatures of crystallisation. The finite shear rate rheology
studies were carried out by performing the shear deformation
simulation using Lees-Edwards boundary conditions and solving the
Langevin equation of motion (see Eqn. 1). The damping coefficient is
chosen such that, after the elementary shear deformation, the system
relaxes, within a accessible time to a potential energy (PE) which is
comparable to the PE we would obtain if we performed a conjugate
gradient minimisation.

We monitor the virial stress as well as the virial pressure evolution
in the system with shear strain to make sure the system reaches a
steady state.  We find that even though the stress seems to have
reached steady state at around $50\%$ strain, the pressure is
evolving. As the pressure saturates at roughly $300\%$, the system
shows a homogeneous flow profile. We shear strain an additional
$200\%$ to obtain good statistics for the observables.  Even though
the velocity profile shows homogeneous flow, we find noticeable
fluctuations in velocity profile, over small strain
windows ($\sim 1\%$). To account for these fluctuations we
compute the local shear rates $\dot{\gamma}_l = d V_x / dz$. The local
rearrangement of particles is characterised by the non-affine mean
square displacement (MSD). To compute non-affine displacement, we
divide the system into slices of thickness $2\sigma$ along the
$z-$direction. Each slice contains approximately
around $4000$ particles. In each of these slice, we compute the shear
frame displacements which are given by
\begin{eqnarray}
	\Delta x_{s} &=& x(\gamma_2) - x(\gamma_1) - V_x(z) \Delta t \\ \nonumber
	\Delta y_{s} &=& y(\gamma_2) - y(\gamma_1),
\end{eqnarray}
\noindent where $\Delta x_s$ and $\Delta x_s$ are the displacements of
the particles in shear frame measured as a function of local strain
difference $\Delta \gamma = \dot{\gamma}_l \Delta t$, $\dot{\gamma}_l$
is the slab specific shear rate measured from the velocity profile,
$x(\gamma_{\alpha})$ and $y(\gamma_{\alpha}$) are the coordinates of
the particles at a particular accumulated shear strain of
$\gamma_{\alpha}$ and $V_x(z) \Delta t$ measures the local accumulated
affine displacement, where $V_x(z)$ is the instantaneous velocity
calculated over the time difference $\Delta t$. The mean square
displacement is computed as $\left < \Delta r^2 \right > =\left <
\Delta x_{s}^2 + \Delta y_{s}^2 \right >$. The MSD as a function of
$\Delta \gamma$ is averaged over different planes along the gradient
direction. The MSD, for a wide range of strain, shows isotropic
behaviour.

\subsection{Displacement correlation function} The displacement correlations \cite{weeks_JPCM_2007}, associated with droplet displacements $S_U$
and mobility (or deviation from average displacements) $S_{\delta U}$
are defined as
\begin{eqnarray}
	S_U(\Delta r, \Delta \gamma) &=& \frac{\left < \vec{U}_i
          \cdot \vec{U}_j \right >}{\left < U^2 \right >} \nonumber
        \\ S_{\delta U}(\Delta r, \Delta \gamma) &=& \frac{\left <
          \delta{U}_i \cdot \delta{U}_j \right >}{\left < \delta U^2
          \right >}
\end{eqnarray}
\noindent where $\vec{U} = U_x \hat{i} + U_y \hat{j}$ is the
displacement vector of a droplet in the shear frame, in the
flow-vorticity plane and $\delta U = \left | U - \left< U \right >
\right |$ is the fluctuation in the displacement of a droplet over the
average displacement, which also known as mobility. The $S_U$ is a
vectorial correlation function which is indicative of directional
correlation between the droplets. This function can be further
decomposed into longitudinal and transverse correlations given by
\begin{eqnarray}
	S_U^L &=& \left < U_i^L U_j^L \right >/\left < U^2 \right > \nonumber \\
	S_U^T &=& \left < \Vec{U}_i^T \cdot \vec{U}_j^T \right >/\left < U^2 \right >
\end{eqnarray}
\noindent where $U_{i,j}^L = \vec{U}_{i,j} \cdot \hat{R}_{ij}$ is the
displacement along the separation vector $\hat{R}_{ij}$ and
$\vec{U}_{i,j}^T = \vec{U}_{i,j} - U_{i,j}^L\hat{R}_{ij}$ is along
perpendicular direction to the separation vector
\cite{weeks_JPCM_2007}. The longitudinal and transverse correlation
function are normalized such that $S_U^L + S_U^T = S_U$.

\end{methods}

\noindent{\bf Supplementary Information} is available in the online
version of the paper.

\noindent{\bf Competing financial interests}\\
\noindent The authors declare no competing financial interests.

\end{document}